\documentstyle[aps,pre,multicol,epsfig]{revtex}
\tighten
\draft
\begin{document}
\preprint{cond-mat/9905014}
\title{Phase Diagram Of The Biham-Middleton-Levine Traffic Model In Three
 Dimensions}
\author{H. F. Chau\footnote{electronic address: hfchau@hkusua.hku.hk} and K. Y.
 Wan}
\address{Department of Physics, University of Hong Kong, Pokfulam Road, Hong
 Kong}
\date{\today}
\maketitle
\begin{abstract}
 We study numerically the behavior of the Biham-Middleton-Levine traffic model
 in three dimensions. Our extensive numerical simulations show that the phase
 diagram for this model in three dimensions is markedly different from that in
 one and two dimensions. In addition to the full speed moving as well as the
 completely jamming phases, whose respective average asymptotic car speeds
 $\langle v\rangle$ equal one and zero, we observe an extensive region of car
 densities $\rho$ with a low but non-zero average asymptotic car speed. The
 transition from this extensive low average asymptotic car speed region to the
 completely jamming region is at least second order. We argue that this low
 speed region is a result of the formation of a spatially-limited-extended
 percolating cluster.  Thus, this low speed phase is present in $n > 3$
 dimensional Biham-Middleton-Levine model as well.
\end{abstract}
\medskip
\begin{multicols}{2}
\section{Introduction}
\label{S:Intro}
 With the ever increasing computational power, simulating traffic in the
 microscopic level by means of cellular automaton becomes a real possibility.
 One of the simplest model for city traffic of this kind is the so-called
 Biham-Middleton-Levine (BML) traffic model \cite{BML}.
\par
 The one-dimensional BML model is simply the elementary binary CA rule~184
 operating on a one-dimensional lattice with periodic boundary condition. The
 asymptotic car speed $\langle v\rangle$ in this one-dimensional model is
 exactly known, and is given by
\begin{equation}
 \langle v\rangle = \left\{ \begin{array}{ll} 1 & \hspace{0.25in} \mbox{if~}
 \rho \leq 1/2, \\ \frac{1}{\rho} - 1 & \hspace{0.25in} \mbox{if~} 1/2 < \rho
 \leq 1, \end{array} \right. \label{E:1-D-v}
\end{equation}
 where $\rho$ is the car density in the system \cite{1-d}. In other words, in
 the one dimensional BML model, traffic jam occurs only when the car density
 $\rho$ is equal to $1\equiv \rho_c^{(1)}$ and all cars move in full speed
 whenever $\rho \leq 1/2$.
\par
 The two-dimensional BML model considers the motions of north- and east-bound
 cars in a two-dimensional square lattice with periodic boundary conditions in
 both the north-south and east-west directions. (We shall give the exact rules
 for the two-dimensional BML model in Section~\ref{S:G_BML}.) Although we lack
 an exact analytical expression for the average asymptotic car speed $\langle v
 \rangle$ as a function of car density $\rho$ in the two-dimensional BML model,
 extensive numerical simulations \cite{BML} as well as mean field theory
 studies \cite{2D_Mean_Field} have been carried out. Their results strongly
 suggest a fluctuation-induced first order phase transition in $\langle v
 \rangle$. Moreover, the average asymptotic car speed is likely to follow
\begin{equation}
 \langle v\rangle = \left\{ \begin{array}{ll} 1 & \hspace{0.25in} \mbox{for~} 0
 \leq \rho < \rho_c^{(2)}, \\ 0 & \hspace{0.25in} \mbox{for~} \rho >
 \rho_c^{(2)}, \end{array} \right. \label{E:2-D-v}
\end{equation}
 where the critical density $\rho^{(2)}_c$ is numerically found to be about
 0.31 \cite{BML} and is analytically proven to be less than 1/2
 \cite{Upper_Bound}. In addition, Tadaki and Kikuchi found a more subtle phase
 transition related to the final jamming pattern. Their numerical study showed
 that jamming patterns for car density $\rho$ less than about 0.52 are very
 well self-organized. On the other hand, when $\rho$ is greater than 0.52, the
 jamming patterns are random \cite{ranjam}.
\par
 Extension of the BML model to higher dimensions can be regarded as a highly
 simplified model for computer network communication in a hypercube. And from
 the physics point of view, it is natural to investigate the phase diagram as
 well as the upper critical dimension of the BML model in higher dimensions. As
 a pioneer study, we report the result of an extensive numerical study of the
 BML model in three dimensions in this paper. We find that the three
 dimensional model has a richer phase diagram than that in one and two
 dimensions. In addition to the fluctuation-induced first order phase
 transition in $\langle v\rangle$, we also observe a low (but non-zero) speed
 phase.
\par
 To begin, we first introduce the higher dimensional generalization of the
 BML model in section~\ref{S:G_BML}. Then, we report our simulation results in
 section~\ref{S:Sim} and present our analysis of results in
 section~\ref{S:Analysis}. Finally, we draw our conclusions in
 section~\ref{S:Conclude}.
\section{The BML Model}
\label{S:G_BML}
 Let us introduce the modified BML model in $n$-dimensions below. Consider an
 $n$-dimensional $N_1 \times N_2 \times\cdots\times N_n$ square lattice with
 periodic boundary conditions. Each lattice site will either contain no car
 (that is, an empty site) or contain exactly one car moving in the $\hat{e}_i$
 direction. We denote $\rho_i$ the density of cars moving along $\hat{e}_i$.
 (That is, $\rho_i$ equals the number of cars moving along the $\hat{e}_i$
 direction divided by the total number of cars in the system.) We denote the
 total car density of the system by $\rho \equiv \sum \rho_i$ and we define the
 car density vector by $\vec{\rho} \equiv (\rho_1,\rho_2,\ldots,\rho_n)$.
 Initially, cars are placed randomly and independently onto the $n$-dimensional
 square lattice according to a pre-determined car density vector $\vec{\rho}$.
\par
 The dynamics of the cars are governed by the following rules. Each
 $\hat{e}_1$-moving car advances one site along the $\hat{e}_1$ direction
 provided that no car blocks its way. Otherwise, that $\hat{e}_1$-moving car
 stays in its present location. Parallel update is taken for all
 $\hat{e}_1$-moving cars. After this, each $\hat{e}_2$-moving car advances
 one site along the $\hat{e}_2$ direction if no car blocks its way. Otherwise,
 that $\hat{e}_2$-moving car stays in its present location. Again, parallel
 updating is used. This process goes on until each $\hat{e}_n$-moving car is
 given a chance to move. This marks the end of one timestep and the above car
 moving process is repeated over and over again.
\par
 At each timestep, the car speed is defined as the ratio of number of cars
 moved to the total number of cars in the lattice system. And the average
 asymptotic car speed $\langle v\rangle$ is defined as the car speed averaged
 over both the cycle time and initial car configurations. Since we are
 interested in the behavior of the system in thermodynamic limit, so we only
 consider the limit when $N_1,N_2,\ldots,N_n$ all tend to infinity while
 keeping the aspect ratio between each side fixed.
\par
 We define the $n$-dimensional BML traffic model to be the one with aspect
 ratio between each of the $n$ sides being fixed to one. That is to say, $N_1 =
 N_2 = \cdots = N_n$. Also, an $n$-dimensional BML traffic model is called
 homogeneous if and only if $\rho_i = \rho_j$ for all $i,j$
 \cite{stat-phy,thesis}. In this paper, we concentrate on the homogeneous
 three-dimensional BML traffic model. So for simplicity, we shall simply call
 it the three-dimensional BML model whenever confusion is not possible. From
 this definition, it is clear that the average asymptotic car speed $\langle v
 \rangle$ is a function of $\vec{\rho}$ only and its value lies between zero
 and one.
\section{Simulation Results Of The Three-Dimensional BML Model}
\label{S:Sim}
 Our simulation is performed on a variety of machines including clusters of
 Sun Sparc and Dec Alpha workstations, various Pentium-based PCs and Power PCs
 as well as the SP2 supercomputer. The estimated total CPU time is about
 300~mips~years. Even so, owing to our CPU time limitations, we can only
 systematically simulate up to a lattice size of $100\times 100\times 100$.
 Nonetheless, we have also simulated for the cases of very small and very large
 car densities up to a lattice size of $1000\times 1000\times 1000$ before
 finally drawing our conclusions.
\begin{figure}[t]
\begin{center}
\epsfig{file=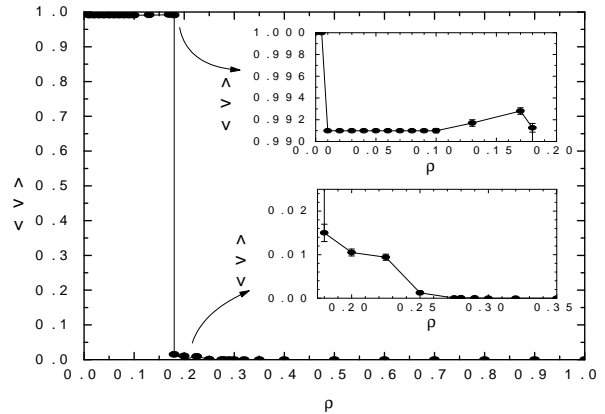,width=9.5cm}
\end{center}
 \caption{The car density $\rho$ vs. average asymptotic car speed $\langle v
  \rangle$ for the three-dimensional BML model on a $100\times 100\times 100$
  lattice. In the two inserted blowups, one clearly observes the drop of
  $\langle v\rangle$ from 1 to $N/(N+1)$ at $\rho = 0.01$ as well as a very
  small bump just below the first critical density $\rho_{c_1}^{(3)} = 0.18\pm
  0.01$. Besides, one also observes that $\langle v\rangle$ reaches zero at the
  critical car density $\rho_{c_2}^{(3)} = 0.32\pm 0.02$.}
 \label{F:speed}
\end{figure}
\par\indent
 Fig.~\ref{F:speed} shows the $\rho$ vs. $\langle v\rangle$ curve for the BML
 model in a $100\times 100\times 100$ lattice. Each data point in the figure
 represents the average asymptotic car speed over an ensemble of random initial
 configurations. For $\rho < 0.1$ as well as $\rho > 0.22$, the value of
 $\langle v\rangle$ is obtained by averaging over $1000$ initial
 configurations. In contrast, $\langle v\rangle$ for $0.1\leq\rho\leq 0.22$ is
 obtained by averaging over only $50$ random initial configuration because the
 long relaxation time prevents us from obtaining more samples.
 Fig.~\ref{F:speed} tells us that $\langle v\rangle = 1$ when the car density
 $\rho\leq 0.005 \approx 1/2N$. We call this region the ``full speed phase''.
 Moreover, recurrent states in the car density region are cycles of period $100
 = N$. (The dependence of various parameters on $N$ here and hereafter are
 based on our simulation results in various lattice sizes up to $1000\times
 1000\times 1000$, including various odd, even and prime values of $N$.) As
 $\rho$ increases to about $0.01 \approx 1/N$, $\langle v\rangle$ begins to
 drop. The recurrent states form cycles with periods several ten times the
 linear lattice size $N$. As the car density $\rho$ reaches about $0.02 \approx
 2/N$, $\langle v\rangle$ drops to the value $N/(N+1) = 100/101 \approx
 0.99099$ and stays constant until $\rho$ reaches about $0.10$. In this car
 density region, recurrent states form cycles of period $101 = N+1$. In other
 words, in the recurrent state, each car in the system will be blocked exactly
 once in each cycle. For the $100\times 100\times 100$ lattice, $\langle v
 \rangle$ is slightly greater than $N/(N+1)$ as $0.10 \leq\rho\leq 0.17$. A
 similar but much smaller bump in $\langle v\rangle$ is also observed in the
 $200\times 200\times 200$ lattice. Hence, we conclude that the bump is due to
 finite size effect of the lattice. Typical recurrent states in this car
 density region are cycles of period $\approx 100N$. Moreover, the typical
 relaxation time for a random configuration in this range of car density
 appears to scale exponentially with $N$. In fact, the long relaxation time
 forbids us from performing systematic simulation with lattice size greater
 than $100\times 100\times 100$. Clearly, the exponentially long relaxation
 time signals a critical slow-down.
\par
 As the car density reaches about $\rho_{c_1}^{(3)} = 0.18\pm 0.01$, $\langle v
 \rangle$ drops abruptly to about $0.015$. Since the asymptotic car speeds in
 all our simulation data are either greater than or equal to $N/(N+1)$ or less
 than $0.03$, we strongly believe that the observed sudden drop in $\langle v
 \rangle$ is a result of a first order phase transition. Interestingly, the
 periods of the recurrent configurations of all these low but non-zero speed
 states are equal to $N$. When we further increase the car density $\rho$,
 $\langle v\rangle$ gradually decreases until it finally reaches zero at
 $\rho^{(3)}_{c_2} = 0.32\pm 0.02$.
\par
 In summary, our simulation tells us that for a finite $N\times N\times N$
 lattice, the system exhibits a non-trivial ``high speed region'' with $\langle
 v\rangle = N/(N+1)$ as well as a non-trivial ``low speed region'' with
 $\langle v\rangle \lesssim 0.03$. Thus, in the thermodynamic limit, the
 three-dimensional BML model has a full speed phase, a low speed phase and a
 completely jammed phase. (Moreover, just like the two-dimensional case, the
 completely jammed phase may further be divided into the self-organized
 jamming and the random jamming regions.) The transition from the full speed to
 the low speed phase is first order in nature and the transition from the low
 speed phase to the completely jammed phase is smooth. That is to say, this
 transition is at least second order.
\section{Analysis Of Our Simulation Results}
\label{S:Analysis}
\subsection{The Full Speed Phase}
\label{SS:Full_Speed}
\begin{figure}[t]
 \begin{center}
 \epsfig{file=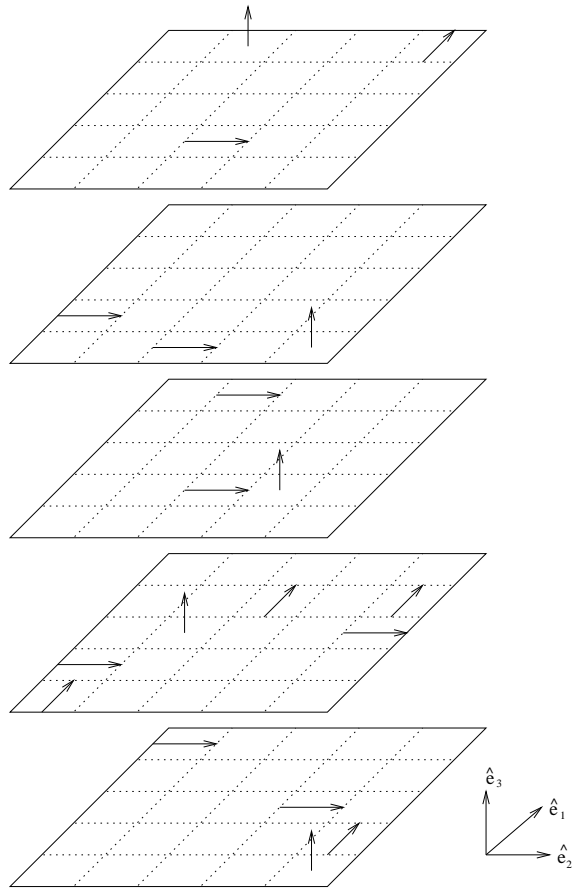,width=7.5cm}
 \end{center}
 \caption{A typical high speed phase recurrent configuration of period six in a
  $5\times 5\times 5$ lattice. Readers may verify that each car in this
  configuration is blocked exactly once per period.}
 \label{F:high_speed}
\end{figure}

 In addition to the systematic trend that the critical car density from the
 high to low speed phase strictly decreases with spatial dimension, we observe
 an interesting feature in the high speed phase of the three-dimensional BML
 model. Unlike the one- and two-dimensional models, the recurrent
 configurations for any finite $N\times N\times N$ lattice in the high speed
 region with $\rho \gtrsim 1/N$ form cycles of period $N+1$. A typical high
 speed recurrent configuration in a $5\times 5\times 5$ lattice is shown in
 Fig.~\ref{F:high_speed} as an illustration. Readers may verify that cars in
 these high speed configurations will be blocked once per cycle period
 \cite{thesis}. Unfortunately, we do not have a good explanation why this
 $\langle v\rangle = N/(N+1)$ recurrent state is preferred over the $\langle v
 \rangle = 1$ recurrent state in three dimensions.
\subsection{The Transition To Low Speed Phase}
\label{SS:Transition}
\begin{figure}[t]
 \begin{center}
 \epsfig{file=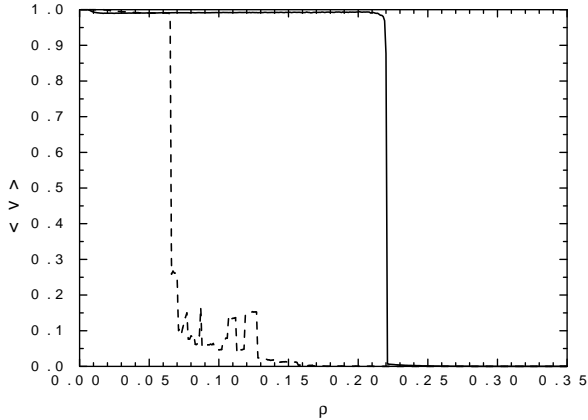,width=9.5cm}
 \end{center}
 \caption{A typical evolution of car density $\rho$ vs. asymptotic car speed
  $\langle v\rangle$ in a $100\times 100\times 100$ lattice when we drive the
  system slowly as discussed in the text. The solid and dash curves represent
  the evolution of the system when cars are slowly added to and removed from
  the system, respectively.}
 \label{F:Hystersis}
\end{figure}

 Since we cannot find any intermediate speed asymptotic configurations in our
 simulation, the transition from the high to low speed phase is likely to be
 first order. To further investigate to the nature of this transition, we drive
 the system by slowly adding cars to or removing cars from the system. That is
 to say, starting from $\rho = 0$, we increase the car density by a fixed small
 amount $\Delta\rho$ by randomly introducing cars to the empty sites in the
 system. And then, we evolve the system until it relaxes to a recurrent state.
 We repeat the process until $\rho$ reaches one. After this, we decrease the
 car density of the system by $\Delta\rho$ by randomly removing cars from the
 system. And then, we evolve the system until it reaches a recurrent state. We
 repeat this process until $\rho$ becomes zero. The $\rho$ vs. $\langle v
 \rangle$ graph obtained in this way on a $100\times 100\times 100$ lattice
 with $\Delta\rho = 0.001$ is shown in Fig.~\ref{F:Hystersis}. Clearly, as we
 slowly increase the car density $\rho$, transition to the low speed phase
 occurs at car density around $0.22$, which is slightly higher than the
 critical car density $\rho_{c_1}^{(3)} \approx 0.18$. More dramatically, as we
 slowly decrease the car density, transition to the high speed phase occurs at
 car density around $0.07$, which is much smaller than the critical car density
 $\rho_{c_1}^{(3)}$. The observed hysteresis loop confirms the hypothesis that
 this is a first order phase transition. And since the only nonlinearity in the
 model comes from the exclusion volume effect, we conclude that the phase
 transition is fluctuation induced.
\subsection{The Low Speed Phase And The Completely Jamming Phase}
\label{SS:Low_Speed}
\begin{figure}[t]
 \begin{center}
 \epsfig{file=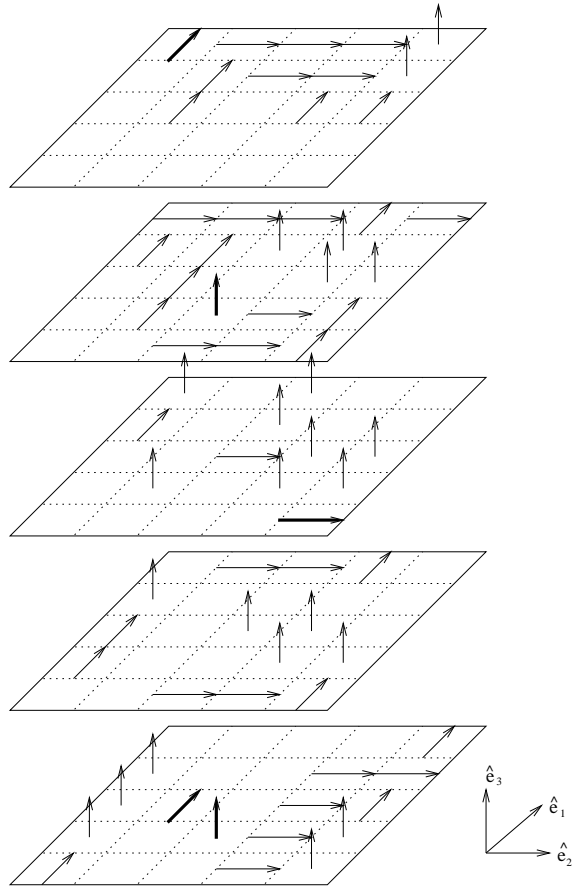,width=7.5cm}
 \end{center}
 \caption{A typical low speed recurrent configuration in a $5\times 5\times 5$
  lattice making up of five full speed cars with $\langle v\rangle = 1/14$. We
  denote the five full speed cars by bold arrows.}
 \label{F:low_speed}
\end{figure}

 Unlike its one- and two-dimensional counterparts, the three-dimensional BML
 model has a low speed phase with $0 < \langle v\rangle \lesssim 0.03$. Similar
 to the completely jamming configurations, we find that the recurrent
 configurations in the low speed phase contain (directed) percolating clusters
 of cars. But unlike the completely jamming configurations, we found a small
 number of residual freely moving cars in the low speed phase. Hence, the
 period of these recurrent states equals the linear system size $N$. And since
 most cars are already jammed by colliding into the percolating cluster, the
 average asymptotic car speed is low. A typical low speed recurrent
 configuration in a $5\times 5\times 5$ lattice is shown in
 Fig.~\ref{F:low_speed} as an illustration \cite{stat-phy}.
\par
 Recall that a percolating backbone is essentially an one-dimensional object.
 Therefore, if the background lattice is one- or two-dimensional, all other
 moving cars will eventually merge into the percolating backbone leading to a
 completely jamming configuration. The situation is completely different when
 the background lattice is at least three-dimensional. In this case, since both
 the trajectories of moving cars and the percolating backbone are essentially
 one-dimensional objects, trajectories of some moving cars may not intersect
 with the percolating cluster. Thus, if the car density is small enough, some
 residual freely moving cars may present in a recurrent configuration giving
 rise to the observed low speed phase.
\par
 As the car density gradually increases in the low speed phase, the size of the
 percolating cluster in the recurrent configuration increases. It becomes more
 and more difficult for the system to accommodate residual freely moving cars.
 Hence, $\langle v\rangle$ gradually decreases until it eventually reaches
 zero. The transition from the low speed phase to the completely jamming phase
 is, therefore, smooth.
\section{Conclusions And Outlook}
\label{S:Conclude}
 In summary, we study the phase diagram of the three-dimensional BML model.
 Similar to the two-dimensional model, a fluctuation-induced first order phase
 transition in asymptotic average car density $\langle v\rangle$ is observed at
 car density $\rho_{c_1}^{(3)} = 0.18\pm 0.01$. We also discover a new low
 speed phase which is absent in one- and two-dimensional models. We argue that
 the existence of this low speed phase is geometrical in nature, and hence this
 phase will exist in higher dimensional BML models as well. It is instructive
 to numerically verify our claims in the four-dimensional model. Unfortunately,
 the amount of computation involved will probably be too high for us at this
 moment. Finally, our simulation suggests that the transition from the low
 speed phase to the completely jamming phase is smooth and occurs at car
 density $\rho_{c_2}^{(3)} = 0.32\pm 0.02$.
\par
 A number of open questions remain. For instance, we do not understand why the
 $\langle v\rangle = N/(N+1)$ high speed states are preferred over the $\langle
 v\rangle = 1$ full speed states in the three-dimensional model. And it is
 meaningful to investigate if this behavior persists in higher dimensions.
\acknowledgments
 We would like to thank P. M. Hui, K.-t. Leung, L. W. Siu and K. K. Yan for
 their useful comments.

\end{multicols}
\end{document}